\documentclass[a4paper,fleqn,usenatbib]{mnras}

\usepackage{newtxtext,newtxmath}

\usepackage[T1]{fontenc}
\usepackage{ae,aecompl}

\usepackage{graphicx}	
\usepackage{amsmath}	
\usepackage{amssymb}	

\usepackage{txfonts}
\usepackage[]{natbib}
\usepackage{lscape}
\usepackage{longtable}
\usepackage{ulem}       
\usepackage[dvipsnames]{xcolor}

\newcommand{\bvo}{$ \left(B-V \right)_0$}
\newcommand{\bv}{$ \left(B-V \right)$}
\newcommand{\ub}{$ \left(U-B \right)$}
\newcommand{\qcol}{$ \left(U-B \right) - 0.72 \mathbf{\cdot} \left(B-V \right)$}

\newcommand{\MV}{\mbox{$M_V$}}

\newcommand{\ebv}{$ E \left( B-V \right)$}
\newcommand{\rv}{$\rm R_V $}

\newcommand{\Msun}{$\rm \,  M_{\odot}$}

\newcommand{\Zsun}{$\rm \,  Z_{\odot}$}

\newcommand{\Teff}{\mbox{$T_{\rm eff}$}}

\newcommand{\vrad}{$v_{\rm rad}$}

\newcommand{\kms}{~$ \rm km \, s^{-1}$}

\newcommand{\Megayr}{Myr}

\newcommand{\ha}{ $H \, {\alpha}$}

\newcommand{\vs}{\textit{vs}}

\title[Early-O stars in Sextans~A]{Ongoing star formation at the outskirts of Sextans~A: \\
  Spectroscopic detection of early-O type stars.
  }
\author[M. Garcia et al.]{
  Miriam Garcia$^{1}$\thanks{E-mail: mgg@cab.inta-csic.es},
  Artemio Herrero$^{2,3}$,
  Francisco Najarro$^{1}$,
  In\'es Camacho$^{2,3}$
  and Marta Lorenzo$^{4}$~ 
\\
$^{1}$Centro de Astrobiolog\'{\i}a, CSIC-INTA. Crtra. de Torrej\'on a Ajalvir km 4.
      28850 Torrej\'on de Ardoz (Madrid), Spain \\
$^{2}$Instituto de Astrof\'{i}sica de Canarias. 
V\'{i}a L\'{a}ctea s/n, E-38200 La Laguna (S.C. Tenerife), Spain\\
$^{3}$Departamento de Astrof\'{i}sica, Universidad de La Laguna, 
Avda. Astrof\'{i}sico Francisco S\'anchez s/n, E-38071 La Laguna (S.C. Tenerife), Spain\\
$^{4}$Universidad Complutense de Madrid, Departamento de F\'{i}sica de la Tierra y Astrof\'{i}sica, E-28040 Madrid, Spain
}

\date{Accepted XXX. Received YYY; in original form ZZZ}

\pubyear{2018}

\begin{document}
\label{firstpage}
\pagerange{\pageref{firstpage}--\pageref{lastpage}}
\maketitle

\begin{abstract}
  %
  With both nebular- and stellar-derived abundances of $\lesssim$1/10\Zsun~ and low foreground extinction,
  Sextans~A is a prime candidate to replace the Small Magellanic Cloud
  as reservoir of metal-poor massive stars and reference to study the metal-poor Universe.
  We report the discovery of two early-O type stars in Sextans~A,
  the earliest O-stars with metallicity $<$1/7\Zsun~ known to date,
  and two additional O9 stars.
  Colour--excess estimates towards individual targets, enabled by spectral typing, manifest
  that internal reddening is neither uniform nor negligible.
  The four targets define a new region of star formation
  far from the optically-brightest centre of the galaxy
  and from its conspicuous  \ion{H}{II} shells, but not devoid of neutral hydrogen.
  In fact, we detect a spatial
  correlation between OB-stars and \ion{H}{I} in Sextans~A and other dIrr's
  that leads us to propose that the neutral phase may be fundamental
  to star formation in low-density environments.
  According to the existing evidence at least two of the targets formed in isolation,
  thus suggestive of an stochastic sampling of the initial mass function
  that would enable low-mass galaxies like Sextans~A to form
  very massive stars.
  The discovery of these four stars provide spatially-resolved, spectroscopic confirmation of
  recent findings suggesting that dwarf galaxies
  can sustain star formation despite the low density of the gas phase.
\end{abstract}




\section{Introduction}
\label{s:intro}

  In a Universe of ever-growing chemical complexity local, metal-poor massive stars
  represent a strong link to the past.
  They hold the key to interpret medium to high redshift starburst galaxies, supernovae and
  $\gamma$--ray bursts,
  and make fundamental ingredients to simulate the evolution of galaxies.
  They are likely involved in the formation of stellar-size black hole binaries
  whose collapse we are now able to detect \textit{via} gravitational waves.
  And finally, they set a proxy to the physics of the very massive, metal-free first stars.

The dwarf irregular galaxy Sextans~A  \citep[$\rm 10^h 11^m 00.^s8 \, -04^d41^m34^s$, dIrr, \textit{aka} DDO~75][]{McC12}
is  interesting in this context
because of its very poor metal content.
The abundances derived from young \ion{H}{II} regions range 12+log(O/H)= 7.49--7.71 \citep{SKH89a,Pi01,Kal05,MLC05}
and the stellar abundances of $\alpha-$ and Fe--group
elements are similarly low ( $ [Fe/H] , [Cr/H] , [Mg/H] \sim -1 $, \citet{KVal04}).
Its poor metal content is also supported by the rather flat UV--continuum
of OB-type stars \citep{Gal17}.
Sextans~A is hence remarkable because
its population is within the grasp of 8-m telescopes \citep[1.33~Mpc,][]{TRS11}
and its $\lesssim$1/10\Zsun~ metallicity is lower than
all other Local Group dwarf galaxies targeted by
studies of massive stars: IC~1613, WLM, NGC~3109, NGC~6822, IC~10 and the Magellanic Clouds.

\begin{figure}
\centering
   \includegraphics[width=0.47\textwidth]{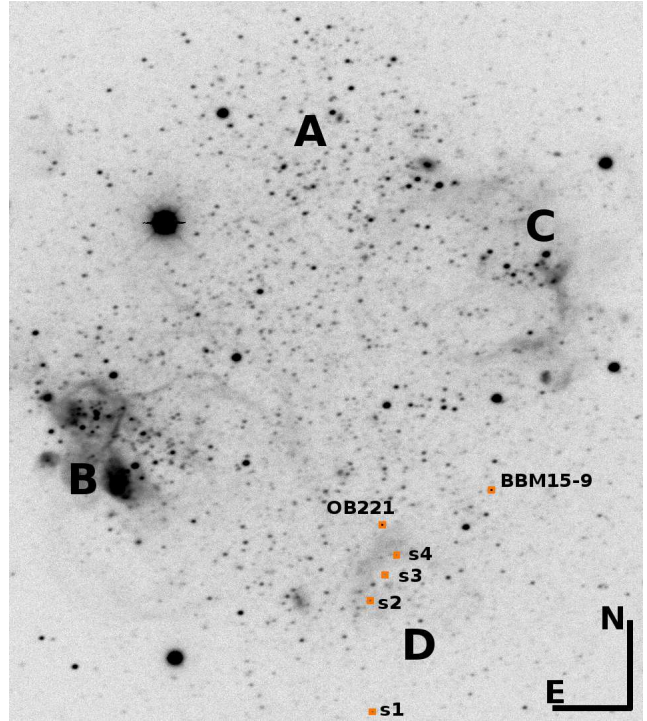}
   \caption{Sextans~A, \ha~ narrow-band image  by \citet{PMal07}.
     The regions where star formation had been previously registered are
     marked A, B and C, and region--D is newly reported in this paper.
     Orange squares mark the location of the O-type stars unveiled
     by our spectroscopic run s1--s4, star OB221 from  \citet{Cal16},
  and the RSG BBM15-9 from \citet{BBM15}.
   \label{F:chart}
}
\end{figure}

Sextans~A has an intriguing squared shape with spectacular bubbles and structures
of ionized hydrogen that evince the presence of hot massive stars.
Ongoing star formation  has been detected in the regions
A, B and C marked in Fig.~\ref{F:chart} \citep{vDPW98,DPal02}.
Region--A would be the oldest one
with 400 million years (\Megayr), and about to exhaust the local gas content and halt star formation.
Region--B overlaps with a vast mass of \ion{H}{I}
and region--C seems to follow the ridge of another over-density of \ion{H}{I}.
They have been forming stars for the past 200~\Megayr~ and 20~\Megayr~ respectively.

Our team confirmed, for the first time, the presence of blue
massive stars in Sextans~A
with long-slit spectroscopic observations \citep{Cal16}.
  In parallel, \citet{BBM14,BBM15} conducted a successful search of
  red supergiant (RSG) candidates from \textit{Spitzer} photometry
  that were subsequently confimed by spectroscopy.
  Massive stars in both OB-type and RSG flavours were found in the three regions of
  star formation, and \citet{Cal16} reported 
  stars as young as 4~\Megayr~ in regions--B and --C.
We concluded that even though region--C may have been activated later on in galactic history, region--B
had been more prolific forming stars and has sustained star formation until
  the present day.

This example illustrates the potential of synergies between spectroscopic surveys
to unveil and characterize massive stars in star-forming galaxies,
with studies of the galaxies themselves.
O-stars and B-supergiants are
H- and very early He-burning massive stars,
younger than $\lesssim$~ 30~\Megayr~ \citep[e.g.][]{M13}.
They pinpoint star formation in both space and time
and can inform current efforts
to understand the nature and fuel of star formation in dwarf galaxies.
On the one hand, studies based on resolved colour--magnitude diagrams and FUV knots indicate
that star formation can proceed despite the low-density
of the neutral-gas phase \citep{McQ12b,Hunt16}.
On the other hand, an important part of the puzzle is missing because
the combination of far distance and low metallicity
prevents the detection of molecular gas in many of these systems.
As a consequence, fundamental questions remain open such as
the apparently more prominent role of \ion{H}{I} over \textsc{H}$\rm _2$~ on star formation in this regime
\citep[e.g.][]{HPBB13},
whether the mechanisms to form stars differ from higher
density environments, and whether these
would translate into a different sampling or slope of the initial mass function.

This paper presents first results of our multi-object spectroscopic programme in Sextans~A,
designed to produce 
the first large sample of resolved, sub-SMC metallicity massive stars.
We focus on a group of very young O-stars that the observations unveiled at the outskirts,
far from the previously considered sites of star formation.
Data and reduction are described in Sect.~\ref{s:redu}.
The new stars and their spectral types are discussed in Sect.~\ref{s:SpT},
and the colour--magnitude diagram of the new star formation region in Sect.~\ref{s:CMD}.
Sect.~\ref{s:discu} explores the implications of the detected stars
in the context of the initial mass function and star formation studies.
Finally, summary and conclusions are provided in 
Sect.~\ref{s:sum}.

\begin{figure*}
\centering
   \includegraphics[width=\textwidth]{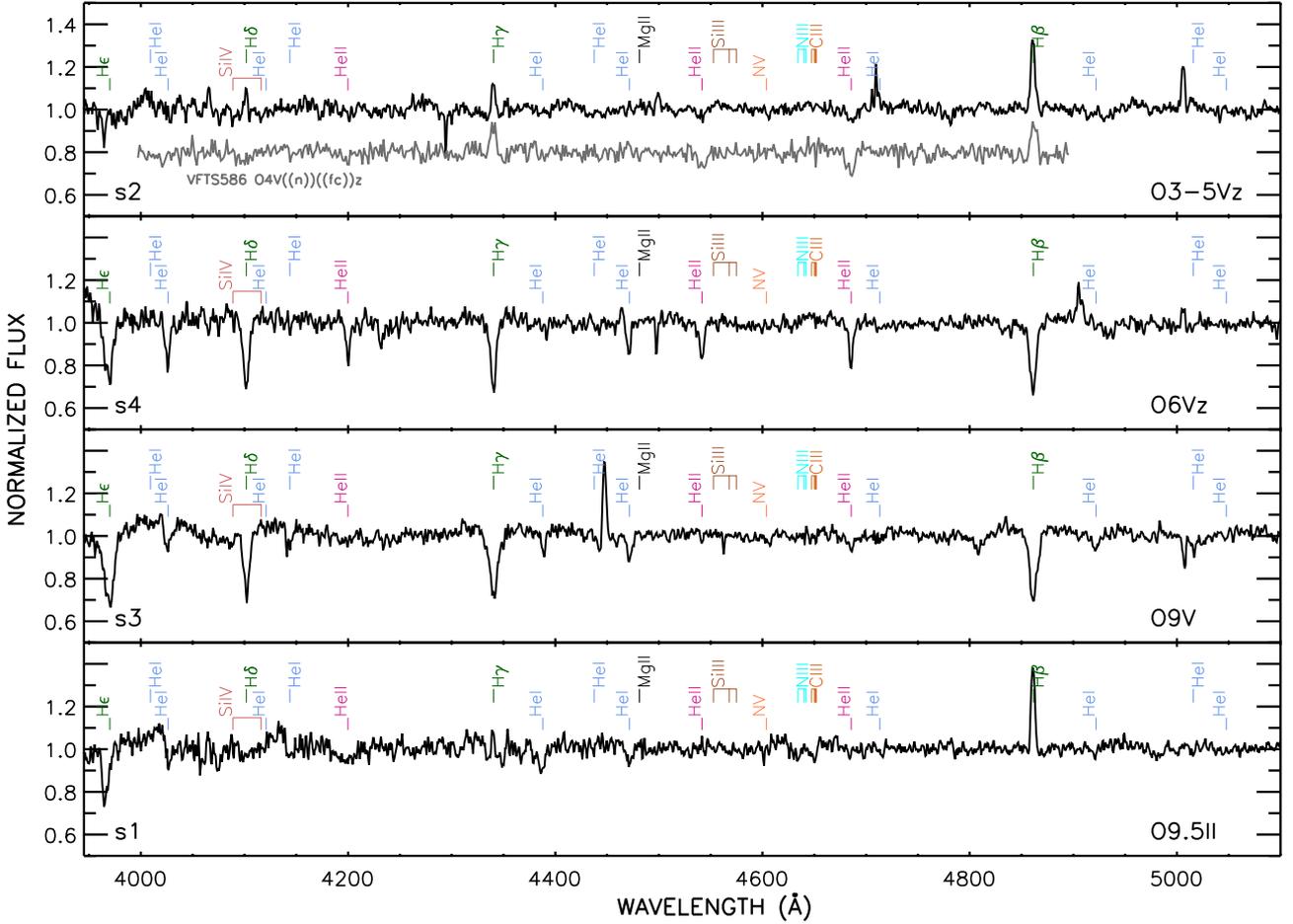}
   \caption{\textit{GTC--OSIRIS} spectra of the discovered O-type stars in Sextans~A.
     Their signal to noise ratio is SNR=30--40.
     All the spectra exhibit one or more spectral lines of \ion{He}{II}.
     The Balmer series is clearly seen in absorption in the spectra of s3 and s4,
     but an incomplete nebular subtraction precludes the identification of the stellar
     component for stars s1 and s2.
     To aid the classification of s2 the top plot includes the \textit{VLT--FLAMES} spectrum of the LMC star VFTS-586,
     degraded to match the spectral quality of our \textit{OSIRIS} run (R=1000 and SNR=40).
}
   \label{F:sp}
\end{figure*}

\section{Observations, data reduction and preliminary analysis}
\label{s:redu}

Data were taken as part of our guaranteed time 
programme GTC3-14AGOS, PI A. Herrero.
The observations consisted on mask multi-object spectroscopy (MOS)
with the Optical System for Imaging and low-Intermediate-Resolution Integrated Spectroscopy (\textit{OSIRIS})
installed at the 10.4-m Gran Telescopio Canarias (\textit{GTC}).
The combination of 1.2~arcsec wide slits and the R2000B VPH
granted resolution $R \sim 1000$~ in the $\sim$4000--5500\AA~ range,
although the actual spectral coverage depends on the slit location within the mask.
The programme was granted 13 hours of gray sky, $<$1.2~arcsec seeing conditions, broken into 1 hour long observing blocks
to accumulate exposure time.

Targets were selected from their optical \ub~ and $Q$=\qcol~ colours,
and ultraviolet \textit{GALEX} photometry,
following the criteria described in \citet{GH13a,Cal16}.
UV sources with $V \leq 21$~ and $Q \leq $--0.8
in \citet{PMal07}'s catalogue were assigned top priority.

The data were reduced with the \textsc{gtcmos}
\textsc{iraf}\footnote{
\textsc{iraf} is distributed by the National Optical Astronomy Observatory, operated by the 
Association of Universities for Research in Astronomy (AURA) under agreement with the National Science Foundation.
}
pipeline developed by Divakara Mayya \citep{GMR16}.
The script transforms the raw CCD images into mosaics, performs
the bias correction and wavelength calibration.
It delivered 13 sky-subtracted, wavelength
calibrated 2-dimensional spectral images, one per observing block. 
The spectra were then extracted with our semi-automatic \textsc{iraf} script
\citep{GH13a} that also performed a second sky subtraction,
and set the spectra to the heliocentric
standard of rest.
The observations did not capture any spectroscopic binary,
as no radial velocity variations were detected among the
13 individual spectra extracted for a given star.
Finally, the individual contributions were coadded
weighing by their signal to noise ratio (SNR), and normalized.

We assigned spectral types following classical criteria, namely
the relative strength of the
lines of different ionization stages of helium
(\ion{He}{I} \vs~ \ion{He}{II}) and silicon (mostly \ion{Si}{III}~ \vs~ \ion{Si}{IV}).
Luminosity class was constrained both by the width of the Balmer lines
and whether \ion{He}{II}4686 is in absorption/emission,
although we note that the latter may be affected by the
low-metallicity of the targets and their weaker winds.

The spectra of the targets subject to this paper are shown in Fig.~\ref{F:sp}
and a detailed discussion of their classification is provided
in the next Section.

\section{A new region of star formation in the South: region--D}
\label{s:SpT}

The spectral classification of the MOS observations revealed the presence of three young O-type dwarf stars
and one O-type bright giant in the South of the galaxy (see Fig.~\ref{F:chart}).
Two of them have the earliest, youngest, spectral types reported in Sextans~A
and it was striking to find them far from the seemingly more active regions B and C.
None the less, three of the O--stars are 
  located near
a faint rim of ionized hydrogen,
a B2.5~supergiant detected by our previous long-slit programme, OB221,
  and RSG number \#9 from \citet[][BBM15-9]{BBM15}. 
The area, that we will name region--D from now on, had been unnoticed by
previous studies targeting the youngest population \citep[e.g.][]{BEH12}.

This is not surprising since region--D is more sparsely populated than regions B and C,
and lacks the classical signs of active star formation: intense UV emission and complex \ha~ structures.
Region--D's unconspicuous appearance could either evince a different star formation regime (Sect.~\ref{ss:SF})
or reflect an incomplete accounting of a heavily reddened population.
In fact, \textit{Herschel} has detected significant amounts of dust in several
locations of the galaxy, some of them close to region--D \citep{SAH14}.
Our own results indicate a non-neglible amount of patchy reddening in the area (Sect.~\ref{s:CMD}).

The membership of the  newly discovered
region--D O--stars to the galaxy is supported by the absolute magnitudes
estimated from spectral types and observed photometry.
Their radial velocities are also consistent with \citet{Sal88}'s radial velocity curve
\citep[contrast against e.g. Fig.~3 from ][]{Cal16},
and with the central \ion{H}{I} velocity $v_{cen}$=324.8km/s \citep{Oal12}.
These data are listed in Table~\ref{T:phot},
together with identification tags and photometry by \citet{PMal07}.

\begin{table*}
  \caption{
    Early-type massive stars confirmed by spectroscopy in Sextans~A:
    Identification codes (ID), spectral types (SpT) and radial
    velocities (\vrad) derived in this work.
    The radial velocities are provided
    in the heliocentric standard of rest, in [\kms].
    They were measured from the Doppler shift experienced by 
    ionized and neutral helium lines, depending on the spectral type, namely
    \ion{He}{II}~4542, \ion{He}{II}~4686, \ion{He}{I+II}~4026, \ion{He}{I}~4144, \ion{He}{I}~4387, \ion{He}{I}~4471 and \ion{He}{I}~4920.
    Identification numbers, coordinates and photometry by \citet{PMal07}
    are also provided.
    Absolute magnitudes \MV~ were calculated 
    using distance modulus DM=25.63$\pm$0.03 \citep{TRS11},
    \bvo~ colours calibrated for their spectral
    types \citep{M98} and adopting \rv=3.1.
  }           
\label{T:phot}      

\centering
\begin{tabular}{l l l l l | l l l l l l }
  \hline
      & This work &                   & \vline~   \citet{PMal07}      &              &             &        &         &          & \vline~       &         \\
ID    &  SpT      & \vrad             & \vline~   ID                  & RA(J2000.0)  & DEC(J200.0) & ~$V$   & \bv     & ~$Q$     & \vline~ \bvo   & \MV     \\
\hline                                                                                                                                            
 s1             & O9.5~II  &  362 $\pm$ 10  & \vline~  J101058.53-044414.4  & 10:10:58.53  & -04:44:14.4 & 20.877 & -0.029  & -1.071   & \vline~ -0.28  & -5.53    \\    
 s2             & O3-5~Vz  &  356 $\pm$ 13  & \vline~  J101058.59-044328.9  & 10:10:58.59  & -04:43:28.9 & 20.804 & -0.095  & -1.120   & \vline~ -0.32  & -5.52    \\  
 s3             & O9~V     &  331 $\pm$ 9   & \vline~  J101058.19-044318.4  & 10:10:58.19  & -04:43:18.4 & 20.803 & -0.247  & -1.005   & \vline~ -0.31  & -5.02    \\  
 s4             & O6~Vz    &  323 $\pm$ 14  & \vline~  J101057.89-044310.2  & 10:10:57.89  & -04:43:10.2 & 20.917 & -0.277  & -1.006   & \vline~ -0.32  & -4.85    \\  
\hline
\hline
\end{tabular}

\end{table*}


\subsection{Comments on targets}

\textbf{s1 (O9.5~II):}
The spectrum lacks strong Balmer lines due to nebular contamination,
and the most clearly detected features belong to \ion{He}{I}.
It shows \ion{He}{II}~4686 but no  \ion{He}{II}~4542,
and the presence of the \ion{Si}{IV}~4089 and \ion{Si}{IV}~4116 lines cannot be
assessed because of the extremely poor SNR at $\lesssim$ 4100\AA.
The \ion{Si}{III}~4552 triplet is present but no \ion{Mg}{II}~4481 is observed,
suggesting O9.5 type.
The weak \ion{He}{II}~4686 absorption and the \ion{He}{II}~4686/\ion{He}{I}~4713 ratio
indicate luminosity class II.

s1 is the farthest sample star from the centre of the galaxy and it is
located in a poorly populated region. 
It is puzzling that it shows strong nebular contamination considering that
no extended structure is seen in \ha~ imaging,
hence ionization must be local
  and could be circumstellar.
We note that it experiences enhanced reddening compared to the other sample stars
(compare e.g. against s3: both have similar spectral type and V--mag$\sim$20.8,
but different colours and luminosity class).
Both pieces of evidence are consistent with s1 located well within the \ion{H}{I} cloud (see Sect.~\ref{ss:SF})
which, in turn,
is additional proof of its Sextans~A membership.
In fact, the neutral hydrogen column--density that would be inferred from
s1's colour excess \ebv=0.251 using \citet{BSD78}'s relations is $N_{HI} $=$ 1.2 \mathbf{\cdot} 10^{21} {\rm cm}^{-2}$,
of the order of the values indicated at its location by the \ion{H}{I} maps \citep{Oal12}.
The presence of s1 is also suggestive that star formation is ongoing in
this region but is undetected because of enhanced extinction.
\textit{Spitzer} does detect additional sources in the surroundings of s1 but
without a thorough analysis no conclusion can be drawn on their ages.

\textbf{s2 (O3--O5~Vz)}
shows wide \ion{He}{II}~4542 and \ion{He}{II}~4686 lines,
and high ionization transitions of silicon,
\ion{Si}{IV}~4089 and \ion{Si}{IV}~4116, whereas the \ion{Si}{III}~4552 triplet is absent.
The lack of \ion{He}{I}~4471 suggests very early spectral type O3
but the combination of nebular contamination, clearly present on the Balmer series,
and poor SNR may be preventing the detection of this line.
A conservative O3--O5 spectral type is assigned.
Because \ion{He}{II}~4542 is weaker than \ion{He}{II}~4686
we assigned Vz luminosity class  \citep[e.g.][]{SSal14}.

The spectrum of s2 is reminiscent of star VFTS-586 (O4~V((n))((fc))z) from the Large Magellanic Cloud (LMC),
which supports its O3--O5~Vz classification.
VFTS-586's high resolution, high SNR \textit{VLT--FLAMES} spectrum shows strong \ion{He}{II} absorptions,
but only narrow nebular emissions  at the \ion{He}{I} transitions
\citep[see e.g.][]{SSSD17}.
In order to match the spectral quality of our observing run, we
degraded the \textit{FLAMES} data to R=1000 and SNR=40.
The overall spectral morphology of VFTS-586 now resembles s2,
both showing \ion{He}{II}~4542 and \ion{He}{II}~4686 as the most
prominent stellar features (Fig.~\ref{F:sp}).

\textbf{s3 (O9~V)}
shows weak spectral lines of \ion{He}{II}.
The \ion{He}{II}~4542/\ion{He}{I}~4471 ratio is compatible with spectral type O9--O9.7
and, because the \ion{Si}{III}~4552 triplet is not detected, O9 spectral type is assigned.
The \ion{He}{II}~4686 absorption is more intense than \ion{He}{II}~4542 and \ion{He}{I}~4713,
which indicates luminosity class III--V. Since Balmer lines are broad,
luminosity class V is adopted.
However, we note that the absolute magnitude \MV=--5.02 is closer to the
calibrated values for O9~III stars (\MV=--5.1) than O9~V (\MV=--4.4).

\textbf{s4 (O6~Vz):}
The \ion{He}{II} lines are strong in absorption, with \ion{He}{II}~4542 slightly
stronger than \ion{He}{I}~4471, indicating O6 type.
The \ion{He}{II}~4200/\ion{He}{I+II}~4026 ratio is concurrent.
Because \ion{He}{II}~4686 is stronger than \ion{He}{II}~4542 and \ion{He}{I}~4471
the assigned luminosity class is Vz.
The absolute magnitude is slightly under-luminous compared
with the calibrated value from Milky Way stars \MV=--5.2.
However, we note that Evans et al. (submitted) have reported that
massive stars in very metal-poor environments
may be up to 0.5~mag fainter than Galactic analogues with the same spectral type.

Two out of the three O~dwarfs reported by this paper have the Vz qualifier,
in line with the trend of increased Vz/V ratios expected
in metal-poorer environments.
\citet{SSal14} used an extensive grid of synthetic models
to study the combination of stellar parameters that could produce the
Vz morphological signature (\ion{He}{II}~4686 absorption stronger than both \ion{He}{II}~4542 and \ion{He}{I}~4471).
They concluded that weak winds are needed to reproduce the Vz characteristics at \Teff$\gtrsim$35000~K,
whereas at lower temperatures no combination of stellar parameters would produce them.
The fact that the earliest O~dwarfs of our sample have the qualifier Vz
  is consistent with this result
and also suggests that the O~Vz stars reflect the weak winds expected at the low-metallicity of Sextans~A.

\section{Colour--magnitude diagram}
\label{s:CMD}

\begin{figure}
\centering
   \includegraphics[width=0.45\textwidth]{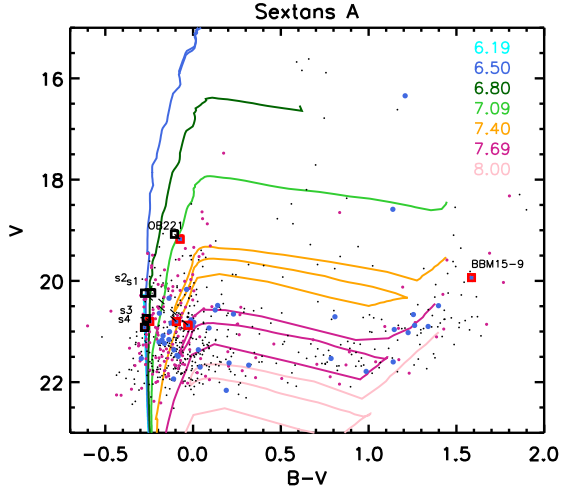}
   \caption{
     Colour--magnitude diagram of Sextans~A.
     \citet{PMal07}'s catalogue for the full galaxy is plotted in black, with region--B stars highlighted in violet
     and region--D stars in blue.  Error bars are omitted for clarity but are shown in Fig.~\ref{F:CMDz}.
     The figure also includes Z=0.001 isochrones by \citet{LJ01}, shifted to account for the distance modulus DM=25.63
     and foreground extinction $ E \left( B-V \right)_{fg}$=0.044 of Sextans~A \citep[][]{TRS11}.
     Their ${\rm log} \left( age \right)$ is colour--coded as indicated in the legend;
     younger isochrones are not included because they would overlap with the ${\rm log} \left( age \right)$=6.19~dex one.
            Red squares mark the observed magnitudes of the programme stars and BBM15-9.
            We also applied an additional reddening correction to the OB-type stars
            from tabulated intrinsic colours 
     and \rv=3.1 (black squares).
}
   \label{F:CMD}
\end{figure}

The colour--magnitude diagram (CMD) of Sextans~A has been
previously analysed in the literature to decipher the galactic 
star formation history (see e.g. Introduction section).
We focus now on the new region
of star formation defined by s1--s4,
that also includes the B2.5 supergiant
OB221           
            and the RSG BBM15-9.
To guide the discussion we will use as reference region--B,
that exhibits more conspicuous signs of star formation,
and has been previously reported to host young stars.

Fig.~\ref{F:CMD} shows Sextans~A's full CMD
built with \citet{PMal07}'s catalogue,
with region--D stars highlighted in blue
and region--B stars in violet.
Most of the stars of both region--B and --D are located in Sextans~A blue plume,
with additional stars with intermediate colours and in the area of red giants/supergiants.
Region--B hosts a comparatively larger number of blue stars
that form a blue envelope to the bulk of the galaxy in the CMD.
Because of the smaller number of stars in region--D, its CMD is scarcely populated
and patchy.
None the less, it also hosts very blue stars and once
their relative numbers are taken into account,
regions--B and --D roughly overlap.
The spectral types of our zone--D sample stars also suggest that
both regions may be similarly young.

The programme stars s1--s4 are found in the bulk of the blue plume of region--D
(Fig.~\ref{F:CMD}).
  BBM15-9 is found in the area of low-mass RSG ($\sim$9\Msun), but its location is not
  reproduced by any of the isochrones. The discrepancy between evolutionary models and observations
  of RSG is a known problem of the field \citep[e.g.][]{DKP13}.
s3 and s4 are among the bluest stars of the galaxy, but s1 and s2 are located
at the red edge of the blue-plume.
In sight of this diagram only OB221,
and perhaps s3 and s4, would have been selected as candidate blue massive stars.

\begin{figure}
\centering
   \includegraphics[width=0.45\textwidth]{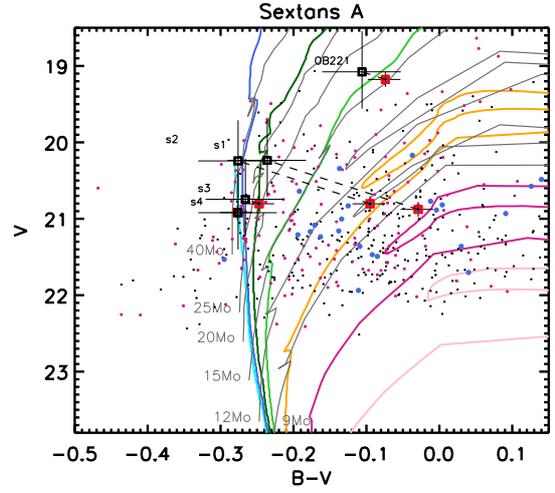}
   \caption{
     Same as Fig.~\ref{F:CMD}, zooming into the main sequence 
       and including evolutionary tracks and error bars.
     s2 is on the track of 40\Msun, s4 and s3 between 25--40\Msun, and s1 at 25\Msun 
}
   \label{F:CMDz}
\end{figure}

The interpretation of the CMD radically changes when reddening is calculated towards each
individual line of sight.
We estimated extinction (\ebv~ and $\rm A_V$) using the observed photometry
and intrinsic colours calibrated for the target's spectral type (listed in Table~\ref{T:phot})
and \rv=3.1.
The so-called \textit{spectroscopic reddening} inherits the uncertainty of the spectral classification ($\pm$2 spectral sub-types) and the tabulated photometry, but the degenerate colors of O-stars in the optical
range minimize the intrinsic error associated to \bvo.
The unknown value of \rv, that varies with dust composition, may play a more prominent role.
A close-up of the CMD is shown in Fig.~\ref{F:CMDz}
  that now includes error bars accounting for
  photometric errors, uncertainty in the distance modulus,
  a conservative error for \rv~ ($\Delta \rm R_V $=2),
  and $\Delta E \left( B-V \right)_{fg}$=0.1 or $\Delta \left(B-V \right)_0$=0.05
  depending on whether foreground or spectroscopic reddening is considered.

Targets s1 and s2 experienced the largest reddening correction and 
are now located at the bluest extent of the blue plume. 
  Even when the error bars are considered,
  the new reddening estimate results in a distinct location in the CMD.
In particular, the updated locus of s2
better matches its spectral type O3--5~Vz.
s1 seems to be much hotter than its assigned O9.5~II spectral type,
although a misclassification seems unlikely in sight of the \ion{He}{II} features.

Reddening would have severely impacted the derived ages for the sample stars.
We have included \citet{LJ01}'s isochrones for Z=0.001 stars ($\sim$ 0.05\Zsun) in Figs.~\ref{F:CMD}~and~\ref{F:CMDz}.
Without any further information on spectral type or reddening, the inferred age of s3 and s4 would have been
${\rm log} \left( age \right)$=6.19--6.80~dex,  
and 7.40--7.69~dex for s1 and s2.        

After the \textit{spectroscopic} reddening correction
s1 overlaps with the 6.80~dex isochrone ($\sim$ 6.3~\Megayr),
rendering a much younger age.
s2, s3 and s4 align around the ${\rm log} \left( age \right)$=6.19--6.5~dex
isochrones (1.5--3.2~\Megayr) and it would be tempting to consider them a coeval OB-association.
However, the population would subtend $\sim$ 25~arcsec (162~pc),
one order of magnitude too loose compared
to, e.g., the  17.5~pc typical sizes of associations in IC~1613 with 3 OB members \citep[Fig.7 from][]{GHC10}.
In sight of Fig.~\ref{F:CMD} all stars of region--D, including  OB221, are younger than
the 7.09dex isochrone (12.3~\Megayr).

The large colour excess of both s1 and s2 also evinces that internal reddening in
Sextans~A is significant  and non-uniform,
  contrary to what is usually assumed for dwarf irregular galaxies.
  Whether this is caused by a patchy distribution of extinction working
  at the cluster scale, or by circumstellar structures surrounding s1 and s2,
  cannot be discerned with current evidence.
  \textit{Spitzer} imaging does not reveal clear point-sources at the
  location of the stars, although this may be a sensitivity issue,
  and \textit{Herschel} data lack the required spatial resolution.

Both s1 and s2 would have been missed by classical CMD cuts
looking for young, massive stars (but not by colour--cuts based
on the reddening-free Q parameter).
The CMD analysis would have yielded masses under 12\Msun~ for both stars,
  while evolutionary tracks assign
25$\rm ^{\texttt{+} 15}_{\texttt{-}10}$\Msun~ and 40$\pm$15\Msun~ 
for the unreddened
locii of s1 and s2, taking into account the error bars (Fig.~\ref{F:CMDz}).
Likewise, the width of region--D's blue plume could be
caused by extinction rather than differences in stellar masses or ages,
rendering the mass of this population heavily underestimated
if internal reddening is neglected.
Integral field spectroscopy covering completely region--D
would at the same time provide stellar masses of all the stars
in the region, and
a measure of extinction towards their line of sight.

\begin{figure}
\centering
   \includegraphics[width=0.45\textwidth]{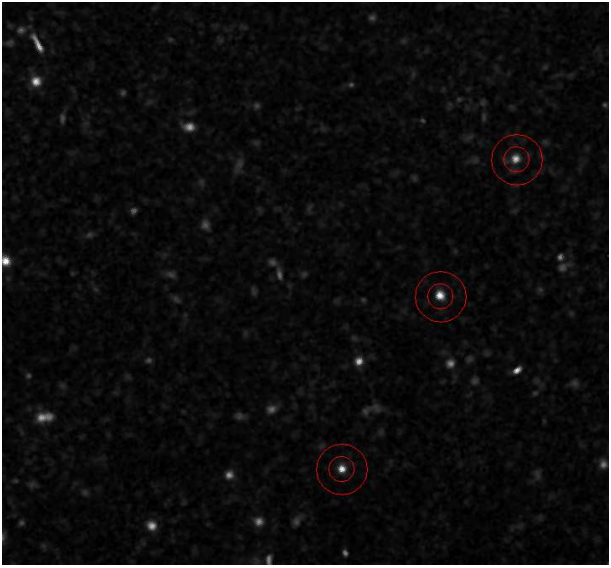}
   \caption{
     \textit{HST-WFPC2-F555W} observations of Sextans~A (programme U2X50205T, PI E. D. Skillman), zooming into s2 (bottom) s3 (middle) and s4 (top).
     s1 is not covered by these or any \textit{HST} observations.
     North is up and East to the left. The circles have 0.77~arcsec and 1.54~arcsec radii,
     corresponding to 5~pc and 10~pc. The stars are isolated except for very faint targets around s4.
}
   \label{F:HST}
\end{figure}

\begin{figure}
\centering
   \includegraphics[width=0.45\textwidth]{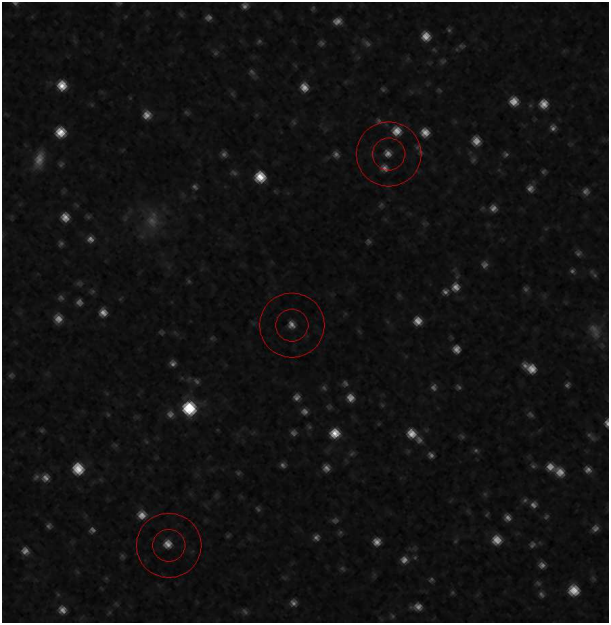}
   \caption{
     \textit{HST-WFC3-F127M} IR imaging (programme icyj11010, PI M. Boyer) around s2--s4 (from bottom to top).
     North is up and East to the left, and the circles have 0.77~arcsec and 1.54~arcsec radii.
     There are two IR-bright stars near s4, but
     s3 and the very early s2 are isolated.
}
   \label{F:HSTIR}
\end{figure}

\section{Discussion}
\label{s:discu}

With ages $\lesssim$~ 10~\Megayr, the stars reported in this
paper provide both spatially and temporally resolved
information on Sextans~A's present day star formation.
In this section we explore what can be learnt from the
location and ages of the stars.

\subsection{Are s1--s4 isolated massive stars? Implications on the initial mass function}

How massive stars form is a long-standing debate 
with two main  scenarios:
competitive accretion, in which massive stars are formed in the same
gravitational well of the whole cluster \citep[e.g.][]{BBC97} and possibly
leading to mergers \citep{Sal12},
and monolithic collapse, in which one single star is
formed from the collapse of one single cloud \citep[e.g.][]{Kral09}.
The occurrence of massive stars in 
isolation is a natural consequence of the latter,
and some examples have been reported
in the SMC \citep{LOW10} and
in the nearby starburst 30~Doradus \citep{BBE12}.

Besides the lack of nearby stars,  \citet{BBE12}
set a number of conditions for the stars to be considered
isolated, including,
 \vrad~ constraints to minimize the chances that the stars
are runaway or binaries,
and the presence of gaseous filaments that
could host star formation and minimize the incidence
of runaways in the plane of the sky.
The lack of a host cluster or OB-association was enforced
in a length-scale of 5~pc, since the
gravitational perturbation of star
formation at longer distances would be negligible
to the forming star.

Lacking multi-epoch observations and high-resolution spectroscopy
the possibility that s1--s4 are binaries or runaway stars cannot be fully discarded,
but their radial velocities are consistent with the \vrad-curve of the galaxy \citep{Sal88,Cal16}
reducing that possibility. We note that the stars are too faint
as to have reliable proper motions registered in \textit{GAIA}-DR2.
Likewise, at the distance of Sextans~A it is not possible 
to identify local filaments of gas although
there is a reservoir of neutral hydrogen at their location (see next Section)
and a faint \ion{H}{II} structure encloses s2, s3 and s4.

Nevertheless, we can check whether the stars are spatially isolated or are part
of a cluster of fainter stars.
We first checked their coordinates against \citet{BBF14}'s list of compact clusters in Sextans~A.
We then examined the archive of the Hubble Space Telescope (\textit{HST}),
looking for observations that would cover them.
Fig.~\ref{F:HST} shows a \textit{WFPC2-F555W} image with s2--s4 
enclosed in a r=5~pc (0.77~arcsec) circle each
to look for a host population,
and circumscribed by a r=10~pc (1.54~arcsec) circle
as a control field.
At the depth of the \textit{HST} observations, 
none of the stars has a similarly bright nearby source within r=5~pc.
There are very faint nearby stars, but none of the them bright enough to be registered even
in \citet{BBF14}'s deep photometric catalogue. 
  No sources are detected near OB221, and BBM15-9 seems to have a faint, very near target at the North.
s1 is at the outskirts of the galaxy and has not been covered by any \textit{HST} observations.
It looks isolated in the ground-based optical and IR images, but
both lack spatial resolution to provide meaningful information for this discussion.

We argue in Sects.~\ref{s:SpT}~and~\ref{ss:SF} that at least 2 of the sample stars are embedded
in neutral hydrogen, and that internal reddening is significant in Sextans~A,
hence it is plausible that the lack of detection of additional stars could be caused by extinction.
\textit{Spitzer} imaging does not reveal a significant dust-enshrouded population in the area
although this could be a matter of sensitivity.
Finally, we examined \textit{HST} near-IR observations 
taken with \textit{WFC3-F127M} \citep{BMcQ17}.
 In this image there is evidence of a small size cluster around
  OB221 and there is no detection of the star close to BBM15-9, which otherwise seems isolated.
  s2--s4 are shown in Fig.~\ref{F:HSTIR}:
there are two IR sources near s4
but no embedded cluster is detected near the target stars and, most importantly,
s3 and the very early s2 are isolated.

Following \citet{LOW10} and \citet{BBE12}, we used
the relation
between the mass of the cluster and the highest stellar mass 
$ M_{max} - M_{cl}$~ from \citet{WKB10,WKPA13}
to estimate the size that a hypothetical cluster hosting s2 (the most massive star of the sample) would have.
\citet{MSH05}'s calibration assigns 47\Msun~ to its spectral type O3--5~Vz.
According to \citet{WKPA13}'s analytical $ M_{max} - M_{cl}$~ relation
the host cluster should have a total mass of $M_{cl} \sim $ 3900\Msun~ and it should have been detected
\citep[e.g. compare against the typical sizes of OB-associations in IC~1613,][]{GHC10}.
  If instead we used the lower-limit we derived for s2 mass from the CMD analysis
  (Sect.~\ref{s:CMD}), the host cluster would be much smaller (725\Msun) but still detectable.

We considered the final possibility that s2 was the product of a stellar merger within a cluster that then would become
an outlier from the $ M_{max} - M_{cl}$~ relation \citep[][]{OK18}.
In the most favourable case, this would require 2 stars of  23\Msun~ each  ($ M_{max}$), for which a cluster of
at least 600\Msun~ is needed. 
Such a cluster should have also been detected.
Morever, the occurrence of the merger is very unlikely if we consider that
only 8 per cent of the clusters simulated by \citet{OK18} are capable
of producing stars with $ 2 \mathbf{\cdot} M_{max}$~ mass via this mechanism.

Current evidence thus indicates that at least
  s2, s3 and perhaps BBM15-9
have formed in isolation
and suggests that monolithic collapse is at work.
In this scenario, star formation does not need to meet the $ M_{max} - M_{cluster}$~ relation
and the  initial mass function (IMF) can be populated randomly \citep[see discussion by ][]{BBE12}.
The stochastic sampling of the IMF has already been proposed to
explain that the linear correlation
between star formation rate (SFR) measured from the UV and \ha~
breaks down in low-mass, low-density galaxies
similar to Sextans~A \citep{TMcNN16}.

If low-mass environments can stochastically populate the initial mass function,
then we could find very massive stars in Sextans~A and
other metal-poor, non-starbursty dwarf irregular galaxies of the Local Group.
Spectroscopically confirmed massive stars have the advantage
of providing precise masses to build the IMF.
A systematic spectroscopic search of massive stars, targeting galaxies
of decreasing metallicity and varying gas masses could shed important light
on to the gas fragmentation properties and
the IMF sampling of these systems.

\begin{figure}
\centering
   \includegraphics[width=0.45\textwidth]{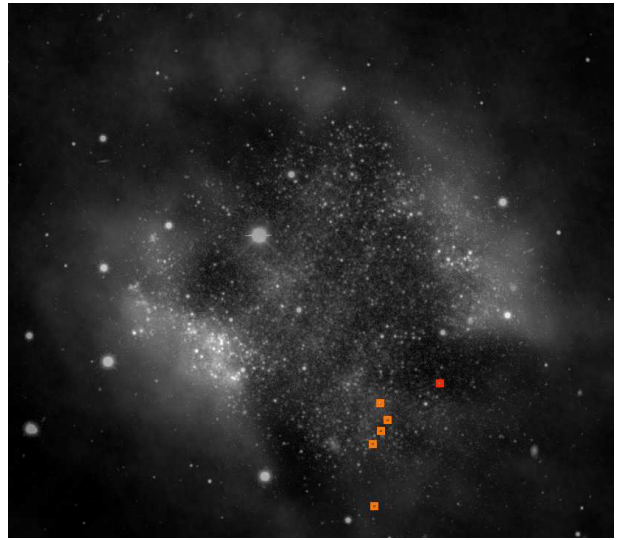}
   \caption{Sextans~A, \textit{LITTLE THINGS} neutral hydrogen map \citep{HFA12}
     overlaid on optical V--band observations \citep{PMal07}.
     The OB-type stars of region--D are highlighted with orange squares.
     We note the comparatively lower, but non-negligible, gas density
     at their location.
       BBM15-9 (red square) is on the edge of the  \ion{H}{I} distribution.
}
   \label{F:chart2}
\end{figure}

\subsection{The mechanisms driving star formation in dwarf galaxies}
\label{ss:SF}

The fact that the earliest O-stars reported to date in Sextans~A
have been found in the outskirts of the galaxy, 
far from its giant \ion{H}{II} shells and without a supporting cluster, was unexpected.
This section discusses whether it is extraordinary that Sextans~A sustains star formation in the outer regions
and what mechanisms could be driving it.

The first radio observations of Sextans~A showed that
the main stellar body is enclosed within an \ion{H}{I} cavity, 
with the youngest regions--B and --C
located  just at the cavity rim or its inner edge.
\citet{vDPW98} proposed that 50~\Megayr~ ago star formation began at the centre of the galaxy
and the cavity was produced by the ensuing supernova explosions
which in turn would induce new star formation at the shocked
outer boundary.  
\citet{DPal02} found older stars in the \ion{H}{I} rim, hence arguing against the outwards propagation of star formation,
but proposed that star formation was confined in the cavity edge.

Higher sensitivity \textit{VLA} data \citep{HFA12,Oal12}
showed that the central part of the galaxy is not totally devoid of gas
($N_{HI} \, \sim 0.5 - 1  \mathbf{\cdot} 10^{21} \, {\rm cm}^{-2}$),
and that \ion{H}{I} rotates as a solid body
with no clear signature of inside--out motion \citep{Oal12,BBF14}.
These observations not only refute the multiple SNe scenario
but also open the possibility of star formation occurring
at additional sites other than the over-densities that delimited
the alleged cavity.

Unfortunately, star formation cannot be traced through
molecular gas in Sextans~A and 
most of the sub-SMC metallicity dwarf galaxies of the Local Group.
They are too far to provide a direct measurement
of the weak signature of cold \textsc{H}$\rm _2$, and their low metallicity prevents using the \textsc{CO} molecule as a proxy.
The situation may improve with \textit{ALMA}, but
at the moment there is only a marginal detection of \textsc{CO} in Sextans~A with \textit{IRAM} \citep{SWZ15}.
A tantalizing alternative is that star formation could proceed directly from \ion{H}{I}
as it has been shown possible at extremely low-metallicities from a theoretical perspective \citep{Kr12}.
This mechanism has also been proposed to explain that a fraction of $z < $0.12 GRBs
are hosted by low-metallicity,
\ion{H}{I}-rich but \textsc{H}$\rm _2$-deficient dwarf galaxies: the galaxy would accrete pristine, cold neutral gas
that could directly feed star formation \citep{MGH15}.

Whether acting as fuel or as proxy of molecular gas,
neutral hydrogen seems a good tracer of star formation in low-density galaxies.
We are finding that the location
of massive stars and \ion{H}{I} is always related.
In IC~1613, as in Sextans~A, O-stars and OB-associations are either found
overlapping the highest
concentrations of \ion{H}{I}, or on the ridges of \ion{H}{I} clouds \citep{GHC10}.
A similar link 
has also been found by \ion{H}{I} surveys on
a significant fraction of star-forming dwarf galaxies  \citep[e.g.][]{TMcNN16,HAE18}
and \citet{HPBB13} proposed that \ion{H}{I} is the dominant phase that regulates star
formation in these systems.

Our sample stars are located in a region that contains \ion{H}{I} (see Fig.~\ref{F:chart2}).
The column density is low $N_{HI} \, \sim 0.5 - 1  \mathbf{\cdot} 10^{21} \, {\rm cm}^{-2}$~ \citep{Oal12}
but close to \citet{S87}'s $1  \mathbf{\cdot} 10^{21} \, {\rm cm}^{-2}$~ threshold for star formation,
and of the order of the specific value for
low-density regions of metal-poor dIrr galaxies proposed by \citet[][1\Msun/$pc^{-2}$]{Hunt16} .
The detection of the O--stars s1--s4 proves that Sextans~A is forming stars in a region with a very
low concentration of gas,
and provides spectroscopic confirmation to similar
findings in the dwarf galaxies targeted by the \textit{LITTLE THINGS} and {SHIELD} \ion{H}{I} surveys \citep{Hunt16,TMcNN16}.

\citet{McQ12b} found that
star formation does not seem spatially concentrated in dwarf galaxies
and that the degree of concentration does not correlate with the peak SFR.
In this context it is plausible that different mechanisms of star formation
work incoherently at a number of galactic locations, 
including the low gas density areas.
In particular, we suggest that star formation in Sextans~A is not driven by molecular
cloud collapse or SNe collect and collapse only. 
These mechanisms could be at work in regions--B and --C,
where the \ion{H}{II} shells hint wind or SN expansion (Fig.~\ref{F:chart})
and localized intense Far-IR emission evinces dust and high concentrations
of molecular gas \citep{SAH14}.

At low density \ion{H}{I} reservoirs like region--D, or at the ridges of \ion{H}{I} distributions, we propose that
internal instabilities or turbulence will break down
the neutral gas clouds and proceed directly to
star formation bypassing the molecular gas phase \citep{Kr12}.
This scenario would be favoured by an irregular or clumpy \ion{H}{I} geometry,
which is indeed detected in the automatic morphological study 
of the \textit{LITTLE THINGS} and \textit{VLA--ANGST} sample by \citet{HPBB13}.

A similar star formation mechanism could also act at the outskirts
of extended UV--disc galaxies,
  a class 
defined by
\textit{GALEX} FUV emission beyond $\rm 3-5 \, D_{25}$~ that
signals star formation in the extremely low-density outer disc
\citep{GdP05,TBB05}.
In fact, 
the location of s1 at $\sim D_{25}$~ \citep[$R_{25}$=2.9~arcmin,][]{McC12}
and other  more remote UV sources make Sextans~A reminiscent of this type of galaxies.
Similarly to what we have detected in Sextans~A, extended UV--disc galaxies are embedded
in large \ion{H}{I} reservoirs, and IR observations do not reveal
an underlying population of low mass or old stars at the sites of 
the FUV complexes \citep[see review by][]{B18}.

Finally, an interesting follow-up question is  whether star formation at different
gas density environments populates the IMF distinctly, or impingnes a different slope.
We will be able to provide some information on this point
once our spectroscopic survey in Sextans~A is complete.

\section{Summary and concluding remarks}
\label{s:sum}

This paper reports the spectroscopic confirmation of massive stars
at the outskirts of Sextans~A.
s2 and s4 are the earliest, most massive, resolved stars confirmed by spectroscopy
in a galaxy with metal content $\lesssim$1/10\Zsun.
Massive stars have
been found in the metal-poorer galaxies
SagDIG \citep{G18} and Leo~P (Evans et al., submitted), but their comparatively poor
data quality prevented fine spectral typing
resulting in poorly constrained masses and ages.
Our sample of stars in Sextans~A is of interest to the community of massive stars,
as new subjects to confront observations with the theoretical
predictions of stellar evolution and wind physics
in the metal-poor regime \citep{K02,Szal15}.

  s1--s4
are only few million year old, thus demonstrating that star formation is ongoing
in a region of comparatively decreased
stellar and gas density 
\citep[$N_{HI} \, \sim 0.5 - 1 \mathbf{\cdot} 10^{21} \, {\rm cm}^{-2}$~
\textit{vs}
the galactic maximum $N_{HI} $=$ 6.1 \mathbf{\cdot} 10^{21} \, {\rm cm}^{-2}$,][]{Oal12}.
However, no direct or indirect signature of molecular
gas has been detected in the area \citep{SAH14,SWZ15}.
Together with the spatial correlation we are finding between
\ion{H}{I} and OB-stars in dwarf irregular galaxies,
this suggests that the neutral phase may be playing a fundamental
role in the process of star formation in low-density environments.

Considering the evidence at hand, two programme stars are isolated and
at least one of them lacks the required
supporting cluster to fully construct the IMF up to its 47\Msun.
Similar isolated massive stars have been found
in the Magellanic Clouds \citep{LOW10,BBE12}.
Our results proof that low-mass dwarf galaxies can not only sustain star formation
but also form very massive stars and, pending deeper IR observations,
they may do so through stochastic sampling of the IMF.
Whether this episode of star formation is inherent to the galaxy \citep{McQCD15}
or environment-induced \citep{DPal02,BBF14}
is left for future work.

This work puts forward new synergies between the communities studying
massive stars and dwarf galaxies.
Direct spectroscopic observations of massive stars, now at reach
in nearby (out to $\sim$1.4~Mpc) dwarf galaxies with 8--10-m telescopes,
provide ideal means 
to study the mechanisms of star formation in these systems.
The joint study of complete, spectroscopic censuses of resolved massive stars,
together with detailed maps of neutral and molecular gas,
will help to establish the connection of star formation and \ion{H}{I},
unravel the relative role played by molecular and neutral gas,
the mechanisms triggering star formation,
whether different mechanisms are at work in different sites of the galaxy,
and whether each of them can populate the IMF distinctly.
At the same time the censuses will enlarge the scarcely populated list of 
  confirmed massive 
stars with metallicity 1/10\Zsun~ or poorer.

The spectroscopic census should be unbiased and complete for two reasons.
Firstly, this paper has demonstrated that massive stars can occur
far from the smoking-gun diagnostics of star formation: ionized gas shells,
intense UV emission and intense Far-IR dust emission.
\citet{McQ12b}  arrived at a similar conclusion 
after studying the galaxy-wide star formation histories of 20 starburst
dwarf galaxies.
Secondly, we have also shown that internal extinction is significant and uneven
in Sextans~A, similarly to IC~1613 \citep{GHV09} and SagDIG \citep{G18}.
The ensuing variable reddening severely
hampers the pre-identification of blue massive stars from classical colour--cuts in the CMD.
In this respect, we would like to remark that 
an unknown amount of massive stars may be missing when galactic mass is calculated
from photometry, and/or their masses underestimated 
because of reddening.
As a consequence, the total galactic stellar mass
may be substantially underestimated with implications
on the computation of the barionic to dark matter ratios.

Our team is already embarked on a vast observational effort
to expose and study the population of massive stars
in Sextans~A using the multi-object spectrographs at the 10-m Gran Telescopio Canarias.
However, the project would greatly benefit from a wide-field
integral-field spectrometer that could comb the whole galaxy, thus avoiding selection biases,
while providing medium resolution spectroscopy at 4000--5000\AA~
to constrain stellar properties.
In this respect, Sextans~A is an ideal target for \textit{BlueMUSE},
an analogue of the powerful \textit{VLT--MUSE} instrument with blue spectral coverage,
currently at concept stage (Bacon et al. 2018, \textit{BlueMUSE} Science Case, Proposal submitted to ESO).

\section{Acknowledgements}

We would like to thank support from MINECO by means of
grants ESP2015-65597-C4-1-R, ESP2017-86582-C4-1-R, AYA2015-68012-C2-1 and SEV2015-0548, 
and from the Gobierno de Canarias under project ProID2017010115.

This paper is based on observations made with the Gran Telescopio Canarias (programme ID GTC3-14AGOS)
installed in the Spanish Observatorio del Roque de los Muchachos 
of the Instituto de Astrof\'{\i}sica de Canarias, on the island of La Palma.
The work has made use of the \textsc{gtcmos} pipeline for the reduction of the
\textit{GTC--OSIRIS} spectroscopic data for which we thank its author Divakara Mayya.
NASA's Astrophysics Data System,
the SIMBAD database \citep{SIMBAD},
and the Aladin Sky Atlas \citep{aladin1,aladin2} were also extensively used.
P. Massey and his team are warmly thanked for publicly sharing their photometric
observations and data of Local Group galaxies.
Finally, we would like to thank our anonymous referee for very constructive comments
and suggestions.












\bsp	
\label{lastpage}
\end{document}